\documentclass[aps,prl,amssymb,amsmath,preprint,floatfix,superscriptaddress]{revtex4-1}
\usepackage{graphicx,color}
\usepackage{epstopdf}
\usepackage{lineno}
\usepackage{booktabs}
\usepackage{soul}
\usepackage{ulem}

\begin{document}
\title{
Laser-induced antiferromagnetic-like resonance in amorphous ferrimagnets
}

\author{S. Mizukami}
\email{shigemi.mizukami.a7@tohoku.ac.jp}
\affiliation{Advanced Institute for Materials Research, Tohoku University, Sendai 980-8577, Japan}
\affiliation{Center for Spintronics Research Network, Tohoku University, Sendai 980-8577, Japan}
\affiliation{Center for Science and Innovation in Spintronics, Tohoku University, Sendai 980-8577, Japan}

\author{Y. Sasaki}
\affiliation{Department of Applied Physics, Graduate School of Engineering, Tohoku University, Sendai 980-8579, Japan}
\affiliation{Advanced Institute for Materials Research, Tohoku University, Sendai 980-8577, Japan}

\author{D.-K. Lee}
\affiliation{Department of Materials Science and Engineering, Korea University, Seoul 02841, Korea}

\author{H. Yoshikawa}
\affiliation{College of Science and Technology, Nihon University, Funabashi, Chiba 274-8501, Japan}

\author{A. Tsukamoto}
\affiliation{College of Science and Technology, Nihon University, Funabashi, Chiba 274-8501, Japan}

\author{K.-J. Lee}
\affiliation{Department of Materials Science and Engineering, Korea University, Seoul 02841, Korea}
\affiliation{KU-KIST Graduate School of Converging Science and Technology, Korea University, Seoul 02841, Korea}

\author{T. Ono}
\affiliation{Institute for Chemical Research, Kyoto University, Gokasho, Uji, Kyoto 611-0011, Japan}

\date{\today}
\begin{abstract}
The magnetization dynamics for ferrimagnets at the angular momentum compensation temperature $T_A$
is believed to be analogous to that for antiferromagnets.
We investigated the pulsed-laser-induced magnetization dynamics in amorphous rare-earth transition-metal ferrimagnet films with a $T_A$ just above room temperature.
For a low pulse fluence, the magnetization precession frequency {\it decreases} as the applied magnetic field increases,
whereas for a higher pulse fluence, it {\it increases} as the applied field increases.
The result was well explained by the left-handed and right-handed precession modes of the antiferromagnetic-like resonance at temperatures below and above $T_A$, respectively, and the data were in agreement with the theoretical simulation.
The study demonstrated the experimental route to achieving antiferromagnetic resonance in ferrimagnets using a pulsed laser.
\end{abstract}

\maketitle
The fundamental research on antiferromagnets started with the classic work of N$\rm \acute{e}$el,
and the spin dynamics in antiferromagnets has been extensively studied in the past \cite{Nagamiya1955},
contributing to the development of the standard theory of antiferromagnetic resonances \cite{Kittel1951}.
Most of the antiferromagnetic resonances have been observed for materials with a lower N$\rm \acute{e}$el temperature using the microwave or infrared technique \cite{Nagamiya1955}.
There has been renewed interest in the utilization of antiferromagnets in spintronic devices beyond the one based on ferromagnets
\cite{Jungwirth2016}.
Recent advances in ultrashort pulse laser and THz wave technology have enabled further exploration of the antiferromagnetic resonance for various antiferromagnets \cite{Kimel2004,Satoh2010,Kampfrath2010},
which is currently still in the early stage of experimental research. 

Herein, we focus on rare-earth (RE) transition-metal (TM) amorphous {\it ferrimagnets}.
Alloys films such as GdFeCo have recently been considered as a prototype of ferrimagnets 
with a perpendicular magnetic easy axis.
They serve as good playgrounds for exploiting the fundamental ultrafast physics \cite{Stanciu2007} as well as spintronic devices \cite{Kaiser2005,Kim2017}.
The RE and TM magnetic moments can be considered as two sublattice magnetic moments coupled antiferromagnetically, leading to a net magnetization tuneable by the composition ratio of RE to TM elements.
The alloys generally have two characteristic temperatures below the Curie temperature: 
the magnetization compensation temperature, $T_M$, at which the two sublattice {\it magnetic moments} are canceled, 
and the angular momentum compensation temperature, $T_A$, at which the two sublattice {\it angular momenta} are canceled.
The existence of $T_A$ in the alloys provides 
a route for exploring the antiferromagnetic-like spin dynamics, even though the alloys are not true antiferromagnets, 
as discussed by Kim {\it et al.} in terms of the domain wall dynamics \cite{Kim2017}.
This means that the antiferromagnetic-like resonance should also be observed in these ferrimagnetic alloys 
at temperatures $T$ near $T_A$. 

The ferromagnetic resonance (FMR) and the exchange modes at $T$ below or around $T_M$ have been well discussed 
in relation to such amorphous RE-TM ferrimagnets and crystalline ferrimagnetic oxides using all-optical pulse laser methods \cite{Stanciu2006,Mekonnen2011,Parchenko2016,Deb2016}.
On the other hand, the antiferromagnetic-like resonance at $T$ around $T_A$ is essentially different from those dynamics
and has not been observed in these alloys.
In this Letter,
we report the observation of antiferromagnetic-like resonance in amorphous GdFeCo ferrimagnets 
at $T$ near $T_A$.
The observed behaviors are consistent with the simple physical pictures described herein and several numerical simulations.
\begin{figure}
\begin{center}
\includegraphics[width=5cm,keepaspectratio,clip]{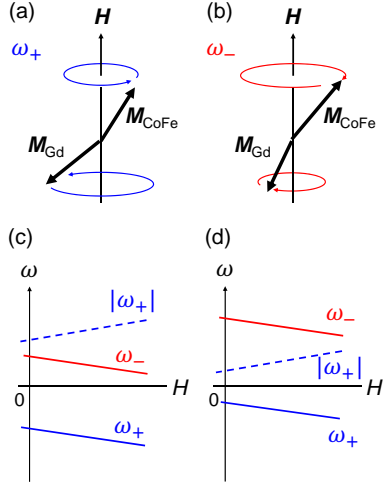}
\caption{
An illustration of the antiferromagnetic-like resonance for ferrimagnets at a temperature $T$ near the angular momentum compensation temperature $T_A$ when the applied magnetic field $H$ is parallel to the magnetic easy axis, {\it i.e.}, the film normal in the present study.
(a) The left-handed mode with the angular frequency $\omega_+$ and (b) the right-handed mode with $\omega_-$.
$\mathbf{M}_{\rm CoFe}$ and $\mathbf{M}_{\rm Gd}$ are the sublattice magnetization vectors for CoFe and Gd, respectively.
The schematic illustration of these mode frequencies {\it vs.} the external magnetic field $H$
at $T$ just below (b) and just above $T_A$ (c).
Dashed lines denote the absolute values of $\omega_+$ {\it vs.} $H$.
}
\end{center}
\end{figure}

The resonance dynamics in ferrimagnetic films with perpendicular magnetic anisotropy (PMA) are discussed based on the coupled Landau-Lifshitz equations for the magnetization vectors of the sublattices $\mathbf{M_{1(2)}}$ \cite{Kittel1951,Geschwind1959}.
The linearized versions of these equations yield the angular frequency $\omega_{\pm}$ for the two modes under an external magnetic field $\mathbf{H}$ applied parallel to the film normal and parallel (antiparallel) to $\mathbf{M_{1(2)}}$ (1: CoFe and 2: Gd) under thermal equilibrium:
\begin{eqnarray}
\omega_\pm &=& \mp \mu_0 \left[\overline{\gamma H_{\rm eff}}^2 + 2 \overline{\gamma H_{\rm eff}} \cdot \overline{\gamma H_{\rm ex}} +[\delta (\gamma H_{\rm ex})]^2 \right]^\frac{1}{2} \\ \nonumber
&-&  \mu_0 [  \overline{\gamma} H + \delta (\gamma H_{\rm k}) + \delta (\gamma H_{\rm ex}) ].
\end{eqnarray}
Here, $\overline{\gamma H_{\rm eff}}=[\gamma_1(H_{\rm k1}+H) + \gamma_2(H_{\rm k2}-H)]/2$, $\overline{\gamma H_{\rm ex}}=(\gamma_1 H_{\rm ex1} + \gamma_2 H_{\rm ex2})/2$, $\delta (\gamma H_{\rm ex}) = (\gamma_1 H_{\rm ex1} - \gamma_2 H_{\rm ex2})/2$, $\delta (\gamma H_{\rm k}) = (\gamma_1 H_{\rm k1} - \gamma_2 H_{\rm k2})/2$, and $\overline{\gamma}=(\gamma_1+\gamma_2)/2$.
$\gamma_{1(2)}$, $H_{\rm k1(2)}$, and $H_{\rm ex1(2)}$ are the absolute values for the gyromagnetic ratio, the effective PMA field, 
and the effective magnetic field of the antiferromagnetic exchange coupling for the magnetization $M_{1(2)}$ of sublattice 1(2), respectively.
$\mu_0$ is the permeability in a vacuum. 
We simplify Eq. (1) to capture the underlying physics, assuming that $H_{\rm k1(2)} >> H$ and that $\delta (\gamma H_{\rm k})$ is negligible.
$\delta (\gamma H_{\rm ex})$ can be rewritten with the mean field coefficient $\lambda$ ($>0$) as
\begin{equation}
\delta (\gamma H_{\rm ex}) = \gamma_1 \gamma_2 \lambda (S_2 - S_1) /2,
\end{equation}
so that it is determined by the difference in the angular momentum density $S_{1(2)}$ ($\equiv M_{1(2)}/ \gamma_{1(2)}$)
for sublattice 1(2).
Since $\delta (\gamma H_{\rm ex})$ may be small at $T$ near $T_A$, 
Eq. (1) may be crudely approximated as follows:
\begin{eqnarray}
\omega_\pm &\approx& \mp \mu_0 \left[\overline{\gamma H_{\rm k}} (\overline{\gamma H_{\rm k}} + 2 \overline{\gamma H_{\rm ex}} ) \right]^\frac{1}{2} \\ \nonumber
&-& \mu_0 [ \overline{\gamma} H + \delta (\gamma H_{\rm ex}) ],
\end{eqnarray}
with $\overline{\gamma H_{\rm k}} = (\gamma_1 H_{\rm k1} + \gamma_2 H_{\rm k2} )/2$.
Equation (3) is a counterpart of the well-known relation of the antiferromagnetic resonance mode in pure antiferromagnets \cite{Kittel1951}.
The $\omega_{+}$ and $\omega_{-}$ modes represent the left-handed and right-handed precession modes, as schematically shown in Figs. 1(a) and 1(b), respectively.
The absolute value of $\omega_{+(-)}$ increases (decreases) as the magnetic field increases, 
which results from the opposite gyration motion 
being similar to the pure antiferromagnetic resonance.
Different from the pure antiferromagnetic resonance,
the correspondence of the respective high and low frequency modes either to the $\omega_{+}$ and $\omega_{-}$ modes or to the $\omega_{-}$ and $\omega_{+}$ modes varies at $T$ smaller or larger than $T_A$, as schematically shown in Figs. 1(c) and 1(d).
This phenomenon occurs because $\delta (\gamma H_{\rm ex})$ in Eq. (3) behaves as a negative or positive offset at $T$ smaller or larger than $T_A$.
This change in the attribution may be the unique characteristic of the antiferromagnetic-like resonance for ferrimagnets and should be experimentally examined.

The sample studied is the 30-nm-thick amorphous thin films of Gd$_{23}$Fe$_{67.4}$Co$_{9.6}$,
which were fabricated on thermally oxidized Si substrates by a magnetron sputtering method.
The 5-nm-thick SiN layers were deposited as a buffer and capping layer.
The film exhibited a net magnetization of 45 kA/m and a perpendicular magnetic anisotropy field of approximately 1 T at room temperature, measured by a vibrational sample magnetometer.
The sample exhibited $T_M = 239$ K and $T_A = 321$ K, evaluated by the anomalous Hall effect and the domain wall velocity measurement for different temperatures, respectively \cite{Kim2017}.
The time-resolved magneto-optical Kerr effect was measured under the ambient temperature
using the all-optical pump-probe setup
with a Ti:Sapphire laser and a regenerative amplifier, the same as that previously reported \cite{Mizukami2010,Mizukami2011,Iihama2014,Mizukami2016}.
The duration, the central wave length, and the repetition rate for the output laser pulse 
in this study were $\sim$120 fs, $\sim$800 nm, and 1 kHz, respectively.
The angle of incidence of the $p$-polarized pump and $s$-polarized probe beams 
were $\sim$3$^\circ$ and $\sim$8$^\circ$, respectively, with respect to the film normal. 
The respective spot sizes for the pump and probe beams, which were focused on the film surface with spatial overlapping, 
were 1.3 and 0.37 mm in diameter.
The maximum magnetic field applied was 2 T, with variable field directions. 
\begin{figure}
\begin{center}
\includegraphics[width=8.5cm,keepaspectratio,clip]{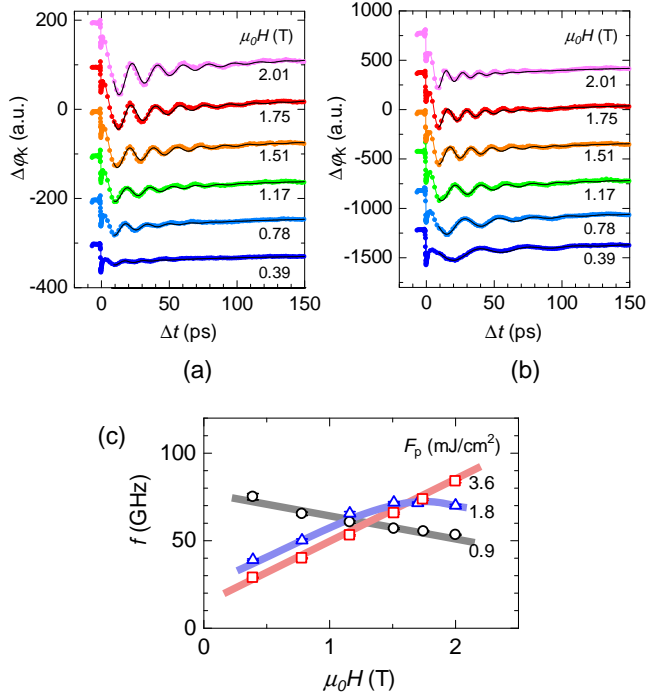}
\caption{
The change in the Kerr rotation angle $\Delta \phi_k$ as a function of the pump-probe delay time $\Delta t$ for different magnetic field strengths $H$ was measured at pump pulse fluences $F_p$ of 0.9 (a) and 3.6 mJ/cm$^2$ (b). 
Both data were collected for a field angle $\theta_H$ of 70$^\circ$ with respect to the film normal. 
Solid curves are fitted to the data. (c) The precession frequency $f$, evaluated from the time-resolved data as a function of $H$, for different $F_p$. The lines and curve are visual guides.
}
\end{center}
\end{figure}

Figures 2(a) and 2(b) show the typical data of the change in the Kerr rotation angle $\Delta \phi_k$ as a function of the pump-probe delay time, 
measured for different applied magnetic fields at pump pulse fluences $F_p$ of 0.9 and 3.6 mJ/cm$^2$, respectively.
The magnetic field angle $\theta_H$ was fixed at 70$^\circ$ from the film normal.
The very rapid changes in the Kerr rotation angle observed at the delay zero 
mainly result from the sub-ps reduction in the normal component of the net magnetization owing to the absorption of the pump pulse.
Subsequent damped oscillations of $\Delta \phi_k$, corresponding to the magnetization precession,
are observed, and their oscillation periods vary with the magnetic field strength.
At a fluence of 0.9 mJ/cm$^2$, the oscillation period becomes longer as the magnetic field strength increases [Fig. 2(a)].
The opposite trend is observed when the fluence is 3.6 mJ/cm$^2$ [Fig. 2(b)].
In addition to the precession clearly visible in the figures, 
additional precession is observed, with much smaller amplitudes and relatively short precession periods (not shown here).
We attribute this mode to a high-frequency branch due to the lift of degeneracy for the $\omega_-$ and $\omega_+$ modes in  ferrimagnets, as discussed earlier.
Since it was very hard to simultaneously fit both the high- and low-frequency modes, 
we analyzed only the low precession frequency mode by fitting the exponentially damped sinusoidal function to the time-resolved data, as shown by the solid curves in Figs. 2(a) and 2(b).
The evaluated frequencies are plotted as a function of the magnetic field in Fig. 2(c) for different pump fluences.
Approximately linear relationships between the frequencies and the magnetic fields are found at fluences of 0.9 and 3.6 mJ/cm$^2$,
regarding which the negative and positive slopes are considered to be those for the $\omega_-$ and $\omega_+$ modes, 
as depicted in Figs. 1(c) and 1(d), respectively.
The mode change as a function of the fluence may stem from the change in the time-averaged sample temperature from below to above $T_A$.
This interpretation is very reasonable since the ambient temperature is just below $T_A = 321$ K
and the sample temperature easily exceeds $T_A$ at the high fluence.
At the intermediate fluence 1.8 mJ/cm$^2$, the frequency falls off near 2 T,
which may be understood as the mode crossing from the $\omega_+$ mode to the $\omega_-$ mode.
Namely, the frequency for the $\omega_-$ mode becomes lower than that for the $\omega_+$ mode at such high field.

Before proceeding further, note the mechanism of the laser-induced magnetization precession.
The primary mechanism of the excitation of these two modes can be attributed to the sudden change in PMA.
This change in the anisotropy functions as the effective torque triggering magnetization precession, 
as discussed in relation to the all-optical FMR \cite{Kampen2002}.
The effective torque works only when the magnetization makes an angle with respect to the magnetic easy direction or plane
and reaches its maximum (zero) when the magnetic field is parallel (perpendicular) to the film plane in the present case.
The magnetization precession amplitudes tend to decrease as the magnetic field angle $\theta_H$ decreases. Hence, it is difficult to observe the dynamics when the applied field is parallel to the film normal in the case depicted in Fig. 1.
Note that the sudden change in the magnetic anisotropy may be caused by the ultrafast demagnetization and its relevant process \cite{Beaurepaire1996}.
More details regarding similar alloys have been discussed for the FMR and exchange modes at $T$ below or near $T_M$ \cite{Stanciu2006,Mekonnen2011}.
However, the mechanism at $T$ near $T_A$ is still unclear, though a discussion on this mechanism is outside the scope of this study.
\begin{figure}
\begin{center}
\includegraphics[width=5cm,keepaspectratio,clip]{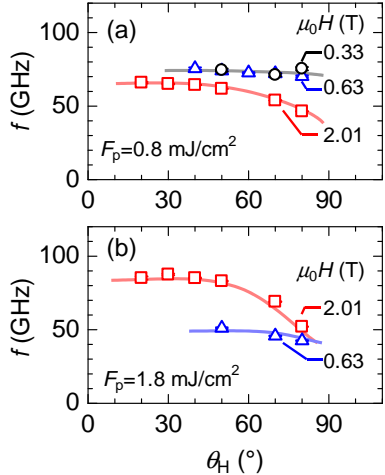}
\caption{
Magnetic field angle $\theta_H$ dependence of the precessional frequency $f$ for low-frequency modes,
extracted from data similarly measured for different field strengths $H$ at a pump laser pulse fluence $F_p$ of 0.8 (a) and 1.8 mJ/cm$^2$ (b). The curves are visual guides.
}
\end{center}
\end{figure}

The magnetic field angle dependence of the precession frequency was also examined to gain insight into the role of PMA in the dynamics observed.
Figures 3(a) and 3(b) show that the precession frequencies increase as the magnetic field angle $\theta_H$ decreases.
This trend is similar to that observed in ferromagnetic films possessing a large PMA, 
such as Co/Pt and CoFeB/MgO multilayers, and ordered alloys films \cite{Mizukami2010,Mizukami2011,Iihama2014,Mizukami2016}.
Thus, this angular dependence may be due to the influence of PMA on the antiferromagnetic-like resonance.
Interestingly, the tendency of the frequency under higher magnetic fields being smaller than that under lower magnetic fields 
is maintained for the different magnetic field angles considered here for a low pump fluence [Fig. 3(a)]; the opposite case is true for a high pump fluence [Fig. 3(b)].

Instead of Eq. (3) describing the simple physics, hereafter we discuss the present dynamics, particularly the angular dependence, 
based on a more realistic micromagnetic simulation using the coupled Landau-Lifshitz-Gilbert equations for two sublattice magnetizations with PMA under various strengths and directions of the external magnetic field \cite{Kim2017}.
The mesh was set to $0.4 \times 20 \times 5$ nm$^3$, 
and the exchange stiffness between the sublattices $A_{\rm CoFe-Gd}$ was taken as 0.04 pJ/m.
To model the dynamics at $T$ below (above) $T_A$, 
we input the following temperature-dependent sublattice magnetization with temperature-independent gyromagnetic ratios for TM and RE: $M_{\rm CoFe} = 615$ (510) kA/m with $\gamma_{\rm CoFe} = 193.6\times10^9$ rad/T$\cdot$s; and $M_{\rm Gd} = 568$ (420) kA/m with $\gamma_{\rm Gd} = 176\times10^9$ rad/T$\cdot$s, respectively.
The difference in the corresponding angular momentum density is $S_{\rm Gd} - S_{\rm CoFe} = 0.506$ $(-2.48)\times10^{-3}$ J$\cdot$s/rad$\cdot$m$^3$.
Note that the magnetization values at $T$ below (above) $T_A$ 
correspond to the values at $T= 312$ (380) K
that were evaluated from the experimental magnetization-temperature curves using the method described in Ref. \onlinecite{Hirata2018}.
The intrinsic PMA constant $K = 0.245$ (0.17) MJ/m$^3$ for both sublattices was assumed for $T$ below (above) $T_A$.
Then, the effective uniaxial PMA constant $K^{\rm eff}_{\rm CoFe (Gd)}$ for the sublattices was given as 
$K^{\rm eff}_{\rm CoFe (Gd)} = K -2 \pi M_{\rm CoFe (Gd)}^2$.
The two-mode precessional dynamics was computed in the time domain for various field strengths and directions, 
where a sublattice-independent Gilbert damping parameter of 0.001 was input.
Then, the mode frequencies were evaluated via the fast Fourier transform.
Note that we also employed a theoretical calculation based on the fully analytic formula for an arbitrary magnetic field and direction, 
with Eq. (1) being a special case; it yielded data approximately equal to those evaluated by simulations with the parameters listed above as the input.

\begin{figure}
\begin{center}
\includegraphics[width=8.5cm,keepaspectratio,clip]{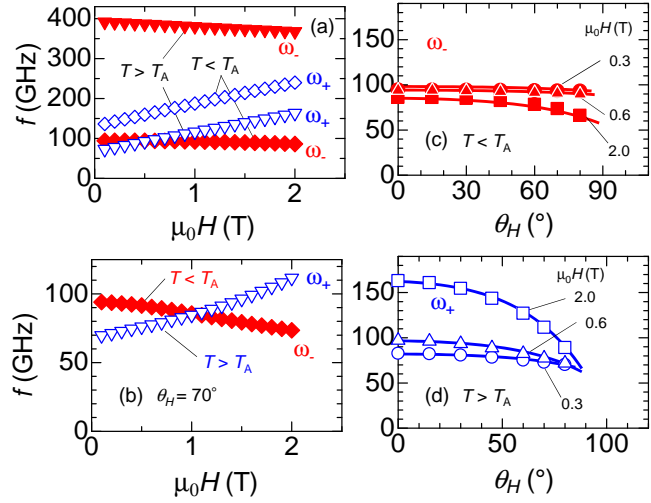}
\caption{
The calculated data of magnetization precession frequency $f$ as a function of the magnetic field $H$ 
with (a) the magnetic field angle $\theta _H= 0 ^\circ$ and (b) $70 ^\circ$.
The data of $\omega_+$ and $\omega_-$ at $T < T_A$ ($T > T_A$) are indicated by the solid and open diamonds (inverse triangles), respectively. 
(b) shows only the modes with lower frequencies.
The calculated data of $f$ {\it vs.} $\theta _H$ for the modes with lower frequencies at $\mu_0 H =$ 0.3 (circle), 0.6 (triangle). and 2.0 T (square)
at (c) $T < T_A$ and (d) $T > T_A$.
The input parameters correspond to the ones at $T$ just below or above $T_A$; see the main text. The curves are visual guides.
}
\end{center}
\end{figure}
Figure 4 displays the computed mode frequencies.
The data reveal a nearly linear relationship between the frequency and the strength of the out-of-plane magnetic field [Fig. 4(a)], in which the high- (low-) frequency mode exhibits a positive (negative) slope in the case of $T$ below $T_A$ (the opposite is true in the case of $T$ above $T_A$), 
being roughly consistent with Eq. (3) and similar to that shown in Figs. 1(c) and 1(d).
The computation also shows that the negative and positive slopes of the $f$ {\it vs.} $H$ data are similarly observed as the low-frequency mode even when the magnetic field angle is 70$^\circ$, as experimentally verified [Fig. 4(b)].
Subsequently, the experimental field-angle variation was verified via a computation with different angles and a fixed magnetic field.
The calculated data for $T$ below and above $T_A$ are shown in Figs. 4(c) and 4(d), 
which well reproduce the experimental tendencies displayed in Figs. 3(a) and 3(b), respectively.
Thus, the simulation satisfactorily supports the conclusion that the dynamics observed is the antiferromagnetic-like resonance mode in ferrimagnets.
Note that the frequency range of $\sim$ 50--150 GHz in Figs. 4(c)-4(d) is also 
roughly consistent with the one experimentally observed ($\sim$ 30--90 GHz).
Additional quantitative investigations should be carried out in future studies, 
which will require a precise evaluation of the high-frequency mode at $T$ near $T_A$.

Finally, we comment on the difference between our work and that of Stanciu {\it et al.} \cite{Stanciu2006}.
In their paper, they discussed the FMR and exchange modes at $T$ near $T_M$ and $T_A$
and stated the following: {\it "When the temperature of the sample approaches the angular momentum compensation point, both frequency and the Gilbert damping parameter of the magnetization precession increase significantly. In addition, the high-frequency exchange mode softens and becomes observable."} 
Meanwhile, the theoretical data shown in Fig. 3 in Ref. \onlinecite{Stanciu2006} indicate 
that the frequency of the FMR mode becomes infinite at $T_A$ owing to the divergence of the effective gyromagnetic ratio at this point.
Our study, however, presents a rather different physical picture at $T_A$. 
The two modes exhibit similar frequencies at $T$ near $T_A$, 
originating from the nature of an antiferromagnetic-like state at this point in ferrimagnets. 
The essential difference between the two modes at $T = T_A$ is the left-handed or right-handed symmetry, 
which was experimentally confirmed with the observation of the opposite response to the applied field. 
This outcome is a very natural consequence of the beautiful symmetry in antiferromagnets, namely, a time-reversal invariant.

In summary,
the pulsed-laser-induced magnetization precessional dynamics in the GdCoFe ferrimagnetic film with $T_A$ just above the ambient temperature was reported.
An inversion of the relation of the gyromagnetic precession frequency with respect to the magnetic field
was clearly observed as the pump laser fluence changed.
This inversion was well explained by the change between the right-handed and left-handed precession modes,
being attributed to the antiferromagnetic-like resonance modes at $T$ below and above $T_A$, under laser-induced heating.
This unique dynamics was also examined for different magnetic field angles,
with all experimental data being consistent with the micromagnetic simulation.
The findings of this study will contribute to development of the physics of the antiferromagnetic spintronics underlying these ferrimagnets.

\begin{acknowledgments}
S.M. thanks CSRN, and Y.S. thanks GP Spin of Tohoku University.
This work was partially supported by KAKENHI (16H03846, 26103002, and 26103004).
D.K.L. and K.J.L. were supported by the National Research Foundation of Korea (2017R1A2B2006119) and the KIST Institutional Program (Project No. 2V05750). 
\end{acknowledgments}


\begin{thebibliography}{99}
\bibitem{Nagamiya1955}
T. Nagamiya, K. Yosida, and R. Kubo, 
Adv. Phys. {\bf 4,} 1 (1955).

\bibitem{Kittel1951}
C. Kittel, 
Phys. Rev. {\bf 82,} 565 (1951).

\bibitem{Jungwirth2016}
T. Jungwirth, X. Marti, P. Wadley, and J. Wunderlich, 
Nat. Nanotechnol. {\bf 11,} 231 (2016).

\bibitem{Kimel2004}
A. V. Kimel, A. Kirilyuk, A. Tsvetkov, R. V. Pisarev, and T. Rasing, 
Nature {\bf 429,} 850 (2004).

\bibitem{Satoh2010}
T. Satoh, S.-J. Cho, R. Iida, T. Shimura, K. Kuroda, H. Ueda, Y. Ueda, B. A. Ivanov, F. Nori, and M. Fiebig, 
Phys. Rev. Lett. {\bf 105,} 077402 (2010).

\bibitem{Kampfrath2010}
T. Kampfrath, A. Sell, G. Klatt, A. Pashkin, S. Mährlein, T. Dekorsy, M. Wolf, M. Fiebig, A. Leitenstorfer, and R. Huber, 
Nat. Photonics {\bf 5,} 31 (2010).

\bibitem{Stanciu2007}
C. D. Stanciu, F. Hansteen, A. V. Kimel, A. Kirilyuk, A. Tsukamoto, A. Itoh, and T. Rasing, 
Phys. Rev. Lett. {\bf 99,} 047601 (2007).

\bibitem{Kaiser2005}
C. Kaiser, A. F. Panchula, and S. S. P. Parkin, Phys. Rev. Lett. {\bf 95,} 1 (2005).

\bibitem{Kim2017}
K.-J. Kim, S. K. Kim, Y. Hirata, S.-H. Oh, T. Tono, D.-H. Kim, T. Okuno, 
W. S. Ham, S. Kim, G. Go, Y. Tserkovnyak, A. Tsukamoto, T. Moriyama, K.-J. Lee, and T. Ono,
Nat. Mater. {\bf 16,} 1187 (2017).

\bibitem{Stanciu2006}
C. D. Stanciu, A. V. Kimel, F. Hansteen, A. Tsukamoto, A. Itoh, A. Kirilyuk, and T. Rasing, 
Phys. Rev. B {\bf 73,} 1 (2006).

\bibitem{Mekonnen2011}
A. Mekonnen, M. Cormier, A. V. Kimel, A. Kirilyuk, A. Hrabec, L. Ranno, and T. Rasing, 
Phys. Rev. Lett. {\bf 107,} 117202 (2011).

\bibitem{Parchenko2016}
S. Parchenko, T. Satoh, I. Yoshimine, F. Stobiecki, A. Maziewski, and A. Stupakiewicz, 
Appl. Phys. Lett. {\bf 108,} 032404 (2016).

\bibitem{Deb2016}
M. Deb, P. Molho, B. Barbara, and J.-Y. Bigot, 
Phys. Rev. B {\bf 94,} 054422 (2016).


\bibitem{Geschwind1959}
S. Geschwind and L. R. Walker, 
J. Appl. Phys. {\bf 30,} S163 (1959).


\bibitem{Mizukami2010}
S. Mizukami, E. P. Sajitha, D. Watanabe, F. Wu, T. Miyazaki, H. Naganuma, M. Oogane, and Y. Ando, 
Appl. Phys. Lett. {\bf 96,} 152502 (2010).

\bibitem{Mizukami2011}
S. Mizukami, F. Wu, A. Sakuma, J. Walowski, D. Watanabe, T. Kubota, X. Zhang, H. Naganuma, M. Oogane, Y. Ando, and T. Miyazaki, 
Phys. Rev. Lett. {\bf 106,} 117201 (2011).

\bibitem{Iihama2014}
S. Iihama, S. Mizukami, H. Naganuma, M. Oogane, Y. Ando, and T. Miyazaki, 
Phys. Rev. B {\bf 89,} 174416 (2014).

\bibitem{Mizukami2016}
S. Mizukami, A. Sugihara, S. Iihama, Y. Sasaki, K. Z. Suzuki, and T. Miyazaki, 
Appl. Phys. Lett. {\bf 108,} 012404 (2016).


\bibitem{Kampen2002}
M. van Kampen, C. Jozsa, J. Kohlhepp, P. LeClair, L. Lagae, W. de Jonge, and B. Koopmans, 
Phys. Rev. Lett. {\bf 88,} 227201 (2002).

\bibitem{Beaurepaire1996}
E. Beaurepaire, J.-C. Merle, A. Daunois, and J.-Y. Bigot, Phys. Rev. Lett. {\bf 76,} 4250 (1996).

\bibitem{Hirata2018}
Y. Hirata, D.-H. Kim, T. Okuno, T. Nishimura, D.-Y. Kim, Y. Futakawa, H. Yoshikawa, A. Tsukamoto, 
K.-J. Kim, S.-B. Choe, and T. Ono, Phys. Rev. B {\bf 97,} 220403(R) (2018).

\end{thebibliography}
\end{document}